\documentclass[preprint,showpacs,showkeys,preprintnumbers,amsmath,amssymb,aps]{revtex4-1}

\usepackage{bm,amssymb,amsmath,mathrsfs}
\usepackage[usenames]{color}
\usepackage[colorlinks]{hyperref}

\usepackage{dcolumn}

\newcommand{\bc}{\begin{center}}
\newcommand{\ec}{\end{center}}
\newcommand{\bit}{\begin{itemize}}
\newcommand{\eit}{\end{itemize}}
\newcommand{\bq}{\begin{equation}}
\newcommand{\eq}{\end{equation}}

\begin{document}

\title{Some notes on beam dynamics due to vertical oscillations in an all-electric storage ring}

\author{S.~R.~Mane}
\email{srmane001@gmail.com}

\affiliation{Convergent Computing Inc., P.~O.~Box 561, Shoreham, NY 11786, USA}

\begin{abstract}
A document has recently been posted on the arXiv \cite{EDMarxivMar2015},
describing analytical formulas and results of particle tracking simulations, 
for precision tests of numerical integration algorithms for an EDM (electric dipole moment) storage ring.
In the context of an all-electric storage ring, the authors cite theoretical formulas by Orlov \cite{OrlovTrento2012}.
However, the reference to Orlov is to a talk at a workshop in 2012,
and is unpublished and difficult for independent researchers to access and validate. 
This note rederives and generalizes some of Orlov's principal results, using a Hamiltonian formalism,
and also corrects some details in both Orlov's note \cite{OrlovTrento2012}
and the arXiv post \cite{EDMarxivMar2015}.

\end{abstract}



\maketitle

\setcounter{equation}{0}

A document has recently been posted on the arXiv \cite{EDMarxivMar2015},
describing analytical formulas and results of particle tracking simulations in both magnetic and electric storage rings. 
The goal is to publish benchmarking formulas for use as 
precision tests of numerical integration algorithms for an EDM (electric dipole moment) storage ring.
In the context of an all-electric storage ring, the authors cite theoretical formulas by Orlov \cite{OrlovTrento2012}.
However, that reference is to an unpublished talk by Orlov at a workshop in 2012,
and is difficult for readers
to access and validate, and is moreover not a peer-reviewed reference.
This note rederives and generalizes some of Orlov's principal results. 
Admittedly this note is also not a peer-reviewed document, but it is publicly accessible and the contents can be independently validated.
I also correct some details in both Orlov's note \cite{OrlovTrento2012} and the arXiv post \cite{EDMarxivMar2015}.

The document \cite{EDMarxivMar2015} presents results for models of both magnetic and electric storage rings.
Only all-electric models will be treated below.
I treat a particle of mass $m$ and charge $e$, with velocity $\bm{v} = \bm{\beta}c$ and Lorentz factor $\gamma=1/\sqrt{1-\beta^2}$,
moving in a prescribed external electrostatic field $\bm{E}$.
The speed of light is set to unity below $c=1$.
The specific model of an all-electric ring treated in \cite{EDMarxivMar2015} is a homogenous weak focusing ring.
(See Section 5 of \cite{EDMarxivMar2015}.)
I employ cylindrical coordinates $(r,\theta,z)$ and the design radius is denoted by $r_0$.
(Similarly the reference values of other quantities are also denoted by a subscript ``0'' e.g.~$\gamma_0$, etc.)
The field index $n$ is defined so that, in the median plane, 
the radial electric field component is $E_r \propto 1/r^{1+n}$.
The two cases studied in \cite{EDMarxivMar2015} 
are that of no vertical focusing (cylindrical capacitor, purely radial electric field, field index $n=0$)
and very weak vertical focusing (field index $0 < n \lll 1$).
The authors in \cite{EDMarxivMar2015} denote the vertical direction by $y$ and employ the notation
$\theta_y$ for what I call $d(z/r_0)/d\theta$, the slope of the vertical motion.

For the case of no vertical focusing, the orbit is a vertical spiral with a constant pitch angle
(i.e.~$dz/d\theta$ or $\theta_y$ is constant).
It is stated in Section 5.1 in \cite{EDMarxivMar2015} that
``Y.~Orlov[15] solved the orbital motion for an electrostatic field
with no focusing. In this case, the estimates for the average
values of $\Delta\gamma/\gamma_0$ and $\Delta r/r_0$ take the following form:''
{\em [i.e.~eqs.~(23) and (24) in \cite{EDMarxivMar2015}]}
\begin{align}
\label{eq:koopdgg0}
\biggl\langle \frac{\Delta\gamma}{\gamma_0}\biggr\rangle &= \phantom{-} \langle\theta_y^2\rangle \,\frac{\gamma_0^2-1}{\gamma_0^2+1} \,,
\\
\label{eq:koopdrr0}
\biggl\langle \frac{\Delta r}{r_0}\biggr\rangle &= -\frac{\langle\theta_y^2\rangle}{2} \,\frac{\gamma_0^2-1}{\gamma_0^2+1} \,.
\end{align}
Note that actually no average $\langle\cdots\rangle$ is required on the values of $\gamma$ and $\Delta r$ in this model.
``Ref. 15'' in the above statement is Orlov's note \cite{OrlovTrento2012}.
In fact Orlov did {\em not} derive the above expressions.
(Curiously, Orlov is a coauthor of the arXiv post \cite{EDMarxivMar2015}.)
Equations \eqref{eq:koopdgg0} and \eqref{eq:koopdrr0}
were derived by Ivan Koop, whose analysis was reproduced in an Appendix 
in a paper I published \cite{ManeSpinDecohAllElec},
with Koop's kind permission \cite{Koopprivcomm}.
See eqs.~(A5) and (A7) in \cite{ManeSpinDecohAllElec}, respectively, for the above expressions.
I have personally verified the correctness of Koop's elegant solution.

For the case of nonzero vertical focusing, the authors state 
(Section 5.2 in \cite{EDMarxivMar2015}, the authors denote the field index by $m$)
``With weak focusing such that $0 < m \lll 1$, the parameters
analytically estimated by Y.~Orlov[15] are given by
Equations 25 and 26 below:
{\em [i.e.~eqs.~(25) and (26) in \cite{EDMarxivMar2015}]}
\begin{align}
\label{eq:orlovdgg0}
\biggl\langle \frac{\Delta\gamma}{\gamma_0}\biggr\rangle &= \phantom{-} 0\,,
\\
\label{eq:orlovdrr0}
\biggl\langle \frac{\Delta r}{r_0}\biggr\rangle &= -\frac12\, \langle\theta_y^2\rangle \,,
\end{align}
which hold for times much larger than the period of vertical oscillations.'' 
Recall Ref.~15 in \cite{EDMarxivMar2015} is Orlov's note \cite{OrlovTrento2012}.
In fact, only eq.~\eqref{eq:orlovdgg0} appears in \cite{OrlovTrento2012}.
The second expression eq.~\eqref{eq:orlovdrr0} 
was derived in a later note by Orlov \cite{OrlovTrentoFollowupJan2013} (also unpublished).
Hence of the four formulas eqs.~\eqref{eq:koopdgg0}--\eqref{eq:orlovdrr0},
only eq.~\eqref{eq:orlovdgg0} appears in Orlov's note \cite{OrlovTrento2012}.

Nevertheless, the expressions in eqs.~\eqref{eq:orlovdgg0} and \eqref{eq:orlovdrr0} are correct and I derive (and generalize) them below.
For contact with the analyses in \cite{EDMarxivMar2015} and \cite{OrlovTrento2012}, the independent variable is the time $t$.
The coordinates are $(r,\theta,z)$, the conjugate momenta are $(p_r,p_\theta,p_z)$ and I define $x=r-r_0$ and $p_x=p_r$.
The model is an all-electric homogenous weak focusing ring, with an electrostatic potential $\Phi(r,z)$.
The Hamiltonian is 
\bq
H = \biggl[\, m^2 + p_r^2 + p_z^2 + \frac{p_\theta^2}{r^2} \,\biggr]^{1/2} + e\Phi(r,z) \,.
\eq
Then $H$ does not depend on $\theta$ hence $p_\theta$ is conserved.
There is no rf cavity, so $H$ is also an integral of the motion.
The equations of motion are 
\begin{subequations}
\begin{align}
\frac{dx}{dt} &= \phantom{-}\frac{\partial H}{\partial p_r} = \frac{p_r}{H-e\Phi} \,,
\\
\frac{dz}{dt} &= \phantom{-}\frac{\partial H}{\partial p_z} = \frac{p_z}{H-e\Phi} \,,
\\
\frac{dp_x}{dt} &= -\frac{\partial H}{\partial r} = \frac{p_\theta^2/r^3}{H-e\Phi} -\frac{\partial(e\Phi)}{\partial r} \,,
\\
\frac{dp_z}{dt} &= -\frac{\partial H}{\partial z} = -\frac{\partial(e\Phi)}{\partial z} \,.
\end{align}
\end{subequations}
For the case $n=0$ (no vertical focusing), the potential is logarithmic
$e\Phi = m\gamma_0\beta_0^2\,\ln(r/r_0)$.
However, the analysis below treats bounded vertical oscillations, where the field index is $n>0$.
The potential is given by a hypergeometric function, and to the required order
\bq
\label{eq:phi}
\begin{split}
e\Phi &= \frac{m\gamma_0\beta_0^2}{n} \biggl\{\,
1 -\frac{r_0^n}{r^n}\,{}_2F_1\Bigl(\frac{n}{2},\frac{n}{2};\frac12;-\frac{z^2}{r^2}\Bigr) \,\biggr\}
\\
&= \frac{m\gamma_0\beta_0^2}{n} \biggl\{\,
1 -\frac{r_0^n}{r^n}\,\biggl[\, 1 -\frac{n^2}{2!}\,\frac{z^2}{r^2} +\cdots \,\biggr] \,\biggr\} 
\\
& \simeq
m\gamma_0\beta_0^2\,\biggl[\,\frac{x}{r_0} - \frac{n+1}{2}\,\frac{x^2}{r_0^2} 
+ \frac{n}{2}\,\frac{z^2}{r_0^2}\biggl(1 -(n+2)\frac{x}{r_0} \biggr)\biggr] \,.
\end{split}
\eq
Define $\omega_0 = \beta_0/r_0$ as the angular revolution frequency.
For the vertical motion, to linear order
\bq
\frac{dp_z}{dt} \simeq -\frac{m\gamma_0\beta_0^2}{r_0^2}\, nz = -p_0 \omega_0\, n\,\frac{z}{r_0} \,.
\eq
Also $dz/dt \simeq p_z/H_0$ so
\bq
\frac{d^2z}{dt^2} \simeq -\omega_0^2\, n\,z \,.
\eq
It is well known that the vertical betatron tune is given by $\nu_z^2 = n$.
We parameterize the vertical betatron oscillations using an amplitude parameter $z_0^\prime$ and an initial phase $\phi_{z0}$
\begin{subequations}
\begin{align}
\frac{z}{r_0} &= \frac{z_0^\prime}{\nu_z}\,\sin(\nu_z\omega_0 t +\phi_{z0}) \,,
\\
z^\prime \equiv \frac{p_z}{p_0} &= z_0^\prime\,\cos(\nu_z\omega_0 t +\phi_{z0}) \,.
\end{align}
\end{subequations}
Next we treat the horizontal motion. 
The model treated in \cite{OrlovTrento2012} is that there are no free radial or longitudinal oscillations.
The radial and longitudinal motions are driven by the coupling to the vertical oscillations.
Normally, to linear order, we say that the horizontal and vertical motions are uncoupled,
but in this analysis we include coupling terms of $O(z^2)$ and $O(p_z^2)$,
e.g.~see the expression for the potential $\Phi$ in eq.~\eqref{eq:phi} above.
This means $x = O(z_0^{\prime 2})$ and $p_x = O(z_0^{\prime 2})$ are of the second order in small quantities.
I also set $H = H_0(1+\Delta H/H_0)$, where $\Delta H/H_0$ is also of the second order in small quantities.
Then
\bq
\begin{split}
\frac{p_\theta^2}{r^2} &= (H_0 + \Delta H - e\Phi)^2 -m^2 -p_x^2 - p_z^2 
\\
&\simeq p_0^2 +2H_0(\Delta H -e\Phi) - p_z^2 
\\
&\simeq p_0^2 - p_z^2  +2p_0^2 \biggl(\frac{\Delta H}{H_0\beta_0^2} -\frac{x}{r_0} -\frac{n}{2}\,\frac{z^2}{r_0^2}\biggr) \,.
\end{split}
\eq
Then $dx/dt \simeq p_x/H_0$ and 
\bq
\label{eq:dpxdt}
\begin{split}
\frac{dp_x}{dt} &\simeq 
\frac{p_\theta^2/r^2}{H -e\Phi} \,\frac{1}{r}
-m\gamma_0\beta_0^2\,\biggl[\, \frac{1}{r_0} -(n+1)\,\frac{x}{r_0^2} -\frac{n(n+2)}{2}\,\frac{z^2}{r_0^3} \,\biggr] 
\\
&\simeq
\frac{p_0^2}{H_0r_0}\,\biggl[\, 1 - \frac{p_z^2}{p_0^2}  
+\frac{2}{\beta_0^2}\frac{\Delta H}{H_0} -\frac{x}{r_0} -\frac{n}{2}\,\frac{z^2}{r_0^2}\biggr) \,\biggr]
\biggl(1 -\frac{\Delta H}{H_0} +\frac{e\Phi}{H_0}\biggr)\biggl(1 -\frac{x}{r_0} \biggr)
\\
&\quad
-p_0\omega_0\,\biggl(1 - (n+1)\,\frac{x}{r_0} -\frac{n(n+2)}{2}\,\frac{z^2}{r_0^2} \biggr) 
\\
&\simeq
p_0\omega_0\, 
\biggl( 1 - \frac{p_z^2}{p_0^2}  -\frac{2x}{r_0} -n\,\frac{z^2}{r_0^2} +\frac{2}{\beta_0^2}\frac{\Delta H}{H_0} \biggr)
\biggl(1 -\frac{\Delta H}{H_0} +\beta_0^2\,\frac{x}{r_0} +\beta_0^2\,\frac{n}{2}\,\frac{z^2}{r_0^2} \biggr)\biggl(1 -\frac{x}{r_0} \biggr)
\\
&\quad
-p_0\omega_0\,\biggl(1 - (n+1)\,\frac{x}{r_0} -\frac{n(n+2)}{2}\,\frac{z^2}{r_0^2} \biggr) 
\\
&\simeq -p_0\omega_0\, \biggl[\, (2-\beta_0^2-n)\,\frac{x}{r_0} -\frac{n(n+\beta_0^2)}{2}\,\frac{z^2}{r_0^2} +\frac{p_z^2}{p_0^2} 
-\frac{2-\beta_0^2}{\beta_0^2}\frac{\Delta H}{H_0} \,\biggr] \,.
\end{split}
\eq
This yields
\bq
\frac{d^2x}{dt^2} \simeq -\omega_0^2\, 
\biggl[\, (2-\beta_0^2-n)\,x -\frac{n(n+\beta_0^2)}{2}\,\frac{z^2}{r_0} +r_0\,\frac{p_z^2}{p_0^2} -\frac{2-\beta_0^2}{\beta_0^2}\frac{\Delta H}{H_0} \,\biggr] \,.
\eq
It is well known that the horizontal betatron tune is given by $\nu_x^2 = 2-\beta_0^2-n$.
Since the radial motion consists of bounded oscillations, one must have $\langle dp_x/dt \rangle = 0$.
We use the result $\langle (p_z/p_0)^2 \rangle = \nu_z^2 \langle (z/r_0)^2 \rangle = n \langle (z/r_0)^2 \rangle$ below.
Hence
\bq
\begin{split}
(2-\beta_0^2-n)\,\frac{\langle x\rangle }{r_0} &= 
\frac{n(n+\beta_0^2)}{2}\,\frac{\langle z^2\rangle}{r_0^2} -\frac{\langle p_z^2\rangle}{p_0^2} 
+\frac{2-\beta_0^2}{\beta_0^2}\frac{\Delta H}{H_0}
\\
&= -(2-\beta_0^2-n)\frac{n}{2}\,\frac{\langle z^2\rangle}{r_0^2} 
+\frac{2-\beta_0^2}{\beta_0^2}\frac{\Delta H}{H_0} \,.
\end{split}
\eq
Hence
\bq
\biggl\langle \frac{x}{r_0}\biggr\rangle 
= -\frac{n}{2}\,\biggl\langle \frac{z^2}{r_0^2} \biggr\rangle
+\frac{2-\beta_0^2}{\nu_x^2\beta_0^2}\frac{\Delta H}{H_0} 
= -\frac{1}{2}\,\biggl\langle \frac{p_z^2}{p_0^2} \biggr\rangle
+\frac{2-\beta_0^2}{\nu_x^2\beta_0^2}\frac{\Delta H}{H_0} \,.
\eq
Equation \eqref{eq:orlovdrr0} and Orlov's result in \cite{OrlovTrentoFollowupJan2013}
are special cases of the above for $\Delta H/H_0 = 0$.
Note also that 
\bq
\biggl\langle \frac{e\Phi}{H_0} \biggr\rangle \simeq
\beta_0^2\,\biggl( 
\biggl\langle \frac{x}{r_0}\biggr\rangle +\frac{n}{2}\,\biggl\langle \frac{z^2}{r_0^2} \biggr\rangle \biggr)
= \frac{2-\beta_0^2}{\nu_x^2}\frac{\Delta H}{H_0} \,.
\eq
It follows that 
\bq
\biggl\langle \frac{\gamma}{\gamma_0} \biggr\rangle 
= \biggl\langle \frac{H - e\Phi}{H_0} \biggr\rangle  
= 1 + \biggl(1 - \frac{2-\beta_0^2}{2-n-\beta_0^2}\biggr)\frac{\Delta H}{H_0}
= 1 - \frac{n}{2-n-\beta_0^2}\frac{\Delta H}{H_0} \,.
\eq
Equation \eqref{eq:orlovdgg0} and Orlov's result in \cite{OrlovTrento2012}
are special cases of the above for $\Delta H/H_0=0$.

\bit
\item
Orlov derived eqs.~\eqref{eq:orlovdgg0} and \eqref{eq:orlovdrr0}
in \cite{OrlovTrento2012} and \cite{OrlovTrentoFollowupJan2013} respectively,
but under the approximation of very weak vertical focusing $0 < n \lll 1$.
(This is also stated in \cite{EDMarxivMar2015}.)
I found that such a restriction is unnecessary:
the above derivation did {\em not} require any conditions on the field index 
other than $n>0$ (so as to have bounded vertical oscillations).
Tracking simulations confirm that eqs.~\eqref{eq:orlovdgg0} and \eqref{eq:orlovdrr0} are valid for arbitrary values $0 < n < 1$.
There may be a caveat that the value of $\nu_z$ should not be rational, 
to avoid orbital resonances, to justify the statistical averages.

\item
As stated by Orlov \cite{OrlovTrento2012}, 
the above results are also valid in the presence of rf and synchrotron oscillations.
We can see this as follows:
in the presence of rf, the only change is that the value of $H$ is not constant 
but the average is $\langle H\rangle= H_0$.
We just substitute $\langle \Delta H/H_0\rangle= 0$ in the above derivation, 
and the above expressions for the averages will follow.
Tracking simulations confirm eqs.~\eqref{eq:orlovdgg0} and \eqref{eq:orlovdrr0} are valid in the presence of rf.

\item
As also stated by Orlov \cite{OrlovTrento2012},
the above derivations assume the motion is driven entirely by the vertical betatron oscillations
(p.~14 in \cite{OrlovTrento2012}):
``Assume the presence of only vertical oscillations and the fields as in eq.~(5). 
The energy in (2) depends only on $y$ and, in accordance with this, \dots''
Also p.~2 in \cite{OrlovTrentoFollowupJan2013}:
``It is taken into account that the
shifts $\delta x$ (of the radius) and $\delta\gamma$ 
are quadratic or of a higher-order effect (in the presence of RF).''
As I explained above, the coupling terms in the equations for the radial motion are of $O(z^2)$ and $O(p_z^2)$.
The same remark applies to the synchrotron oscillations.

\item
It is stated after eqs.~(25) and (26) in \cite{EDMarxivMar2015}
(i.e.~eqs.~\eqref{eq:orlovdgg0} and \eqref{eq:orlovdrr0} above)
``Note that these values depend only on the pitch angle, not on the ring geometry, \dots''
It is not clear that eqs.~\eqref{eq:orlovdgg0} and \eqref{eq:orlovdrr0} do not depend on the ring geometry.
Orlov stated (p.~15 in \cite{OrlovTrento2012})
``Taking into account that $(dy/d\tau)^2 / (\beta\gamma)_0^2 = \vartheta_0^2 = m (y/R)^2$, on the average, \dots''
Here $\tau$ is the proper time and $R$ is the design ring radius.
In terms of my notation, this states that
$\langle z^{\prime 2}\rangle = n \langle (z/r_0)^2 \rangle = \vartheta_0^2$.
Such a result is valid only in a homogenous weak focusing ring.
In general, in a model with bends and straight sections (e.g.~drift spaces),
the vertical betatron oscillations must be parameterized using
Twiss functions and the Courant-Snyder invariant must appear in the formulas,
and also $\nu_z^2 \ne n$.
Hence it is not proved that eqs.~\eqref{eq:orlovdgg0} and \eqref{eq:orlovdrr0} do not depend on the ring geometry.

\eit

It should be possible to employ the above formalism also for the case of no vertical focusing, i.e.~$n=0$.
The answer is instructive. 
The solution for the vertical motion in this case is
\bq
\frac{z}{r_0} = z_0^\prime\omega_0 t \,,\qquad
\frac{p_z}{p_0} = z_0^\prime \,.
\eq
The potential is independent of $z$ so we set $n(z/r_0)^2 = 0$. 
We set $\Delta H/H_0=0$ for now (this is important).
Then setting $dp_x/dt = 0$ in eq.~\eqref{eq:dpxdt} 
(and recall that no average $\langle\cdots\rangle$ is required on the values of $x$, etc.) 
yields
\bq
\label{eq:xr0}
\frac{x}{r_0} 
= -\frac{1}{2-\beta_0^2}\frac{p_z^2}{p_0^2} 
= -z_0^{\prime 2}\,\frac{\gamma_0^2}{\gamma_0^2+1} \,.
\eq
This is {\em not} the same as Koop's result in eq.~\eqref{eq:koopdrr0}.
The reason is that $H \ne H_0$ in Koop's derivation. 
Using eq.~\eqref{eq:koopdgg0} and \eqref{eq:koopdrr0} and a logarithmic potential 
and $\ln(r/r_0) \simeq \Delta r/r_0$ yields
\bq
\frac{H}{H_0} = \frac{m\gamma + e\Phi}{H_0}
\simeq 1 + \frac{\Delta\gamma}{\gamma_0} + \beta_0^2\,\frac{\Delta r}{r_0}
= 1 + z_0^{\prime 2}\,\frac{\gamma_0^2-1}{\gamma_0^2+1} 
-\beta_0^2\frac{z_0^{\prime 2}}{2}\,\frac{\gamma_0^2-1}{\gamma_0^2+1}
\\
= 1 + \frac{\beta_0^2z_0^{\prime 2}}{2} \,.
\eq
It is well known that for $n=0$, the centripetal condition for a circular (spiral) orbit 
depends only on the kinetic energy and not on the orbit radius.
The orbit radius is determined by the potential energy.
Koop treats an orbit which is synchronous with the reference particle, and the consequence is that $H\ne H_0$.

However, the value of $\Delta\gamma/\gamma_0$ for a spiral orbit does {\em not} depend on the potential energy,
hence the above formalism should yield eq.~\eqref{eq:koopdgg0}.
Let us verify this. Using eq.~\eqref{eq:xr0},
\bq
\frac{\Delta\gamma}{\gamma_0} = \frac{\Delta H}{H_0} -\frac{\Delta(e\Phi)}{H_0}
= -\beta_0^2\,\ln\Bigl(1+\frac{x}{r_0}\Bigr)
\simeq - \beta_0^2\,\frac{x}{r_0}
= z_0^{\prime 2}\,\frac{\gamma_0^2-1}{\gamma_0^2+1} \,.
\eq
This agrees with eq.~\eqref{eq:koopdgg0}, as required.

In Fig.~9 in \cite{EDMarxivMar2015}, the authors claim close agreement 
of their numerical tracking simulation results with
eq.~\eqref{eq:koopdgg0} (top graph) and eq.~\eqref{eq:koopdrr0} (bottom graph). 
Also in Section 5.2 in \cite{EDMarxivMar2015}, the authors state
``There is an apparent gap between Equations 23 and 24 and Equations 25 and 26 in the limit as $m\to0$. 
The transition between focusing and no focusing can exist since the latter formulas hold only for averages 
over times much larger than the period of vertical oscillations.''
Here the field index is denoted by $m$ and
``Equations 23 and 24'' (resp.~25 and 26) are the formulas without (resp.~with) vertical focusing.
However, the situation is more subtle.
The formulas with vertical focusing (eqs.~\eqref{eq:orlovdgg0} and \eqref{eq:orlovdrr0})
are derived with $\langle H\rangle = H_0$.
There is no cognizance in \cite{EDMarxivMar2015} that for a model with no vertical focusing,
only the value of $\Delta\gamma/\gamma_0$ is uniquely determined
(and the expression in eq.~\eqref{eq:koopdgg0} is independent of the value of $\Delta H/H_0$), 
while the value of $\Delta r/r_0$ is arbitrary and is determined by the potential energy 
(and $H \ne H_0$ in eq.~\eqref{eq:koopdrr0}).
In fact for a model without vertical focusing, the exact value of the radius of a spiral orbit is
\bq
\frac{r}{r_0} = e^{(H - m\gamma)/(H_0\beta_0^2)} = e^{(\Delta H/H_0 - \Delta\gamma/\gamma_0)/\beta_0^2} \,.
\eq
Note that $\Delta H/H_0$ and $\Delta \gamma/\gamma_0$ are {\em not} required to be small.

Also in Orlov's derivations in \cite{OrlovTrento2012} and \cite{OrlovTrentoFollowupJan2013},
the orbit path length $L$ is not changed from the reference value (i.e.~$L = L_0$ after one turn),
if the motion is driven by vertical oscillations, i.e.~there is vertical focusing.
This is {\em not} the case when there is no vertical focusing
and the orbit is synchronous with the reference particle,
which is the case in Koop's derivation of eq.~\eqref{eq:koopdrr0}.
The path length after one turn is
\bq
\frac{L}{L_0} = \Bigl(1 + \frac{\Delta r}{r_0}\Bigr)\,\sqrt{1 + z^{\prime 2}}
\simeq 1 + \frac{\Delta r}{r_0} +\frac{z^{\prime 2}}{2}
\simeq 1 -z_0^{\prime 2}\,\frac{\gamma_0^2-1}{\gamma_0^2+1} +\frac{z_0^{\prime 2}}{2}
= 1 +z_0^{\prime 2}\,\frac{3-\gamma_0^2}{\gamma_0^2+1} \,.
\eq
This fact does not seem to be noted in \cite{EDMarxivMar2015}, nor in Orlov's analyses
\cite{OrlovTrento2012,OrlovTrentoFollowupJan2013}.


\end{document}